\documentclass[a4paper]{article}

\usepackage{INTERSPEECH2022}
\usepackage{pbox}
\usepackage{verbatim}
\usepackage{hyperref}

\title{Comparison of Speech Representations for the MOS Prediction System}
\name{
Aki Kunikoshi, 
Jaebok Kim, 
Wonsuk Jun and K\r{a}re Sj\"{o}lander
}
\address{
  ReadSpeaker
}
\email{\{aki.kunikoshi,jaebok.kim,wonsuk.jun,kare.sjolander\}@readspeaker.com}

\begin{document}

\maketitle

%%===========
%% abstract 
%%===========
\begin{abstract}
Automatic methods to predict Mean Opinion Score (MOS) of listeners have been researched to assure the quality of Text-to-Speech systems. 
Many previous studies focus on architectural advances (e.g. MBNet, LDNet, etc.) to capture relations between spectral features and MOS in a more effective way and achieved high accuracy. 
However, the optimal representation in terms of generalization capability still largely remains unknown. 
To this end, we compare the performance of Self-Supervised Learning (SSL) features obtained by the wav2vec framework to that of spectral features such as magnitude of spectrogram and melspectrogram.
%Moreover, we propose to combine the SSL features and spectral features prediction task, such as melspectrogram or F0.
Moreover, we propose to combine the SSL features and features which we believe to retain essential information to the automatic MOS 
to compensate each other for their drawbacks. 
We conduct comprehensive experiments on a large-scale listening test corpus collected from past Blizzard and Voice Conversion Challenges. 
We found that the wav2vec feature set showed the best generalization even though the given ground-truth was not always reliable. 
Furthermore, we found that the combinations performed the best and analyzed how they bridged the gap between spectral and the wav2vec feature sets.
\end{abstract}
\noindent\textbf{Index Terms}: speech quality evaluation, Mean Opinion Score, MOS prediction, LDNet, SSL-MOS

%%==============
%% Introduction
%%==============
\section{Introduction}
Text-to-Speech (TTS) has been widely adopted in various applications as the quality of synthesis is getting close to that of human speech. 
To measure the quality, Mean Opinion Score (MOS) is often used as a part of subjective evaluations. 
Subjects are asked to rate speech samples from 1 (bad) to 5 (excellent) regarding certain criteria, such as naturalness or similarity to the target voice. 
 
Although the MOS-based evaluation is a gold standard, collecting reliable scores is still challenging. 
Rating is largely affected by the contexts of listening tests, e.g. materials and subjects~\cite{sslmos}. 
Hence, multiple raters are often employed and the average per sample is calculated as ground-truth. 
However, subjective evaluations are still costly and remain as one of major challenges in the field. 
To this end, the task of automatic MOS prediction has been gaining attention in the community.

Training a reliable MOS prediction system requires large-scale corpora e.g. the Blizzard challenge (BC)~\cite{bc2008,bc-vcc2020} and or the voice conversion challenge (VCC)~\cite{vcc2016,vcc2018} where speech recordings and their corresponding MOS are collected. 
While early studies relied on hand-crafted features and simple statistical models~\cite{bc2012}, recent studies proposed complex architectures, e.g. MOSNet~\cite{mosnet}, MetricNet~\cite{metricnet}, MBNet~\cite{mbnet}, and LDNet~\cite{ldnet}, and achieved high accuracy. 
In particular, LDNet~\cite{ldnet} predicted system-level MOS with higher than 0.98 of Spearsmans rank correlation coeffieient (SRCC) in the experiment with VCC2018 data~\cite{vcc2018}.

However, the generalization of these state-of-the-art methods are still questionable. 
When the distribution of test materials is substantially different from that of training materials, the performance significantly drops. 
Since each listening test has own unique contexts, the challenge of generalization is inevitable ~\cite{sslmos}. 

Self-Supervised Learning (SSL) is actively being researched in the field of Machine Learning and is well-known for its strong generalization capability ~\cite{ssl}. 
Especially, SSL using speech or language modalities has demonstrated the state-of-the-art performance on the various tasks such as automatic speech recognition, speaker identification, and language understanding. 
In particular, the wav2vec~\cite{wav2vec,wav2vec2}-based approach showed strong generalization in the task of MOS prediction as well ~\cite{sslmos}. 
While SSL-based approach achieved promising results, there is still room for the improvement and we still raise questions. 
What makes the gap between SSL and non-SSL features such as magnitude of spectrogram and melspectrogram? 
In particular, which circumstance hurts the performance of SSL features? 
How can we bridge the gap? 

For the sake of answering these, we compare the performance of various representations and conducted in-depth analysis in the context of generalization. 
Especially, we employ LDNet as a baseline architecture that deals with variations of ratings and investigate how the reliability of ratings affects the generalization of each representation.

This paper is structured as follows. 
We firstly introduce related studies in Section \ref{sec:relatedwork}. 
Next, we describe LDNet, our baseline system in Section \ref{sec:ldnet}. 
We explain experimental settings and data in \ref{sec:experiment}.
The results will be reported and discussed in Section~\ref{sec:result}, and concluded in Section~\ref{sec:conclusion}.

%%===============
%% Related Work
%%===============
\section{Related Work}\label{sec:relatedwork}
One of the early end-to-end speech objective assessment models is MOSNet~\cite{mosnet}. 
MOSNet uses magnitude of spectrogram and predicts utterance-level MOS. 
Since MOSNet uses the average MOS per utterance and discards each rating per judge in the training phase, bias of each judge and the variation will not be modelled. 
To address this issue, MBNet was proposed~\cite{mbnet}. 
MBNet consists of mean and bias subnets to leverage the individual judge scores of each utterance. 
Few months later, Huang et al. made improvement to the architecture of MBNet and proposed LDNet~\cite{ldnet}. 

While most of the state-of-the-art methods use spectral features e.g. magnitude of spectrogram, adopting SSL features for the task has been recently getting attention due to its strong generalization proven in various speech and language-related tasks. 
For example, wav2vec is a SSL framework that encodes raw speech data via multiple convolutional layers to obtain abstract and latent representations~\cite{wav2vec,wav2vec2}. 
The use of the wav2vec features was investigated to shed a light on the generalization capability for the task of MOS prediction~\cite{sslmos}. 
In particular, wav2vec 2.0~\cite{wav2vec2} was used as an input representation, a simple architecture of mean pooling layer to the input followed by a linear output layer was proposed. 
This simple method already significantly outperforms MOSNet while the use of individual scores of each sentence was not yet leveraged.

In this work, we investigate generalization of various speech representations including wav2vec, magnitude of spectrogram, melspectrogram, and F0. 
We adopt LDNet as a baseline system so that we could leverage the use of individual scores and analyse how the representations interact with the variance of scores.

\begin{comment}
- How SSL features achieve better generalization?
- Do we have sufficient understanding of how SSL features interact with variance of individual scores?
\end{comment}

%%========
%% LDNet
%%========
\section{LDNet}\label{sec:ldnet}
In this section, we briefly describe Listener Dependent Network (LDNet)~\cite{ldnet} which we adopt as a baseline system in our experiments. 

Assume we have a MOS dataset $\mathcal{D}$ containing $N$ speech samples. Each sample $x_i$ has $m$ scores ( listener-dependent scores, hereinafter LD scores) $\{s_i^1,...,s_i^m\}$ rated by a set of listeners $\{l_i^1,...,l_i^m\}$. 
The mean score is denoted as $\bar{s}_i$. MOSNet \cite{mosnet} aims at finding a model $f$ that predicts the $\bar{s}_i$ given a speech sample. 
In the training phase, the MOSNet minimizes a mean square loss regarding the mean score of each sample:
%% eq: MOSNet loss
\begin{equation}
\mathcal{L} = MSE(f(x_i), \bar{s}_i)
\end{equation}
During inference, given an input speech $x$, the trained model is directly used to make the prediction.
%% eq: MOSNet inference
\begin{equation}
\hat{s} = f(x)
\end{equation}

\begin{figure}[t]
    \centering
    \includegraphics[width=0.8\linewidth]{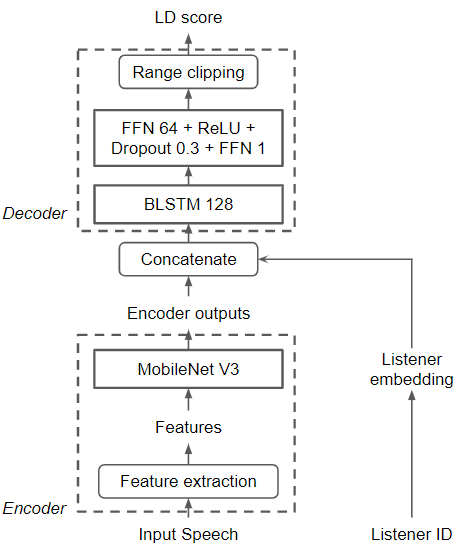}
 \caption{Illustration of the LDNet model architecture.}
 \label{fig:ldnet}
\end{figure}

On the other hand, LDNet \cite{ldnet}, generalized form of MBNet \cite{mbnet}, aims to predict the LD score given the input speech and the listener ID.
LDNet consists of an encoder and a decoder, depicted as figure \ref{fig:ldnet}.
%% fig: LDNet
During the training, LDNet minimizes a LD loss as follows:
%% eq: LDNet loss
\begin{equation}
\mathcal{L} = MSE(f(x_i, l_i^j), s_i^j)
\end{equation}
here,
\begin{equation}
f = Decoder(Encoder(x), l)
\end{equation}

% \subsection{Inference}
% \label{ss:inference}
% LDNet has three inference mode: 
% \begin{enumerate}
% \item all-listeners inference: calculate the prediction of each training listener and average over them.
% \item utilizing the mean score with MeanNet: during the training, train a MeanNet which takes encoder output as input and predict the mean score. At the inference, the MeanNet is used instead of the decoder.
% \item utilizing the mean score with a mean listener: extend
% the training set by adding one virtual ”mean listener” whose score is the mean score. The prediction by the mean listener is assumed to be the final prediction.
% \end{enumerate}

%%=======================
%% Experimental Settings
%%=======================
\section{Experimental settings}\label{sec:experiment}
We conducted experiments to investigate the optimal speech representation for the task of automatic MOS prediction.
As a criteria to compare the performance, we chose the system-level Spearman’s rank correlation coefficient (SRCC) that is used as the main evaluation criteria of VoiceMOS challenge~\cite{voicemos}.

%% result table
%% As two column floats come at the earliest on the page after they appear in the source we need to move the location.
%https://tex.stackexchange.com/questions/110374/how-to-share-a-one-column-figure-with-two-column-text-on-same-page
\begin{table*}[t]
    \caption{Feature Comparison}
    \centering
    \begin{tabular}{l|ccc|ccc|ccc|ccc}
        \toprule
        {}
        & \multicolumn{6}{c|}{Validation data}
        & \multicolumn{6}{c}{Test data}\\        
        \cline{2-13}
        {}
        & \multicolumn{3}{c|}{Utterance level}
        & \multicolumn{3}{c|}{System level}
        & \multicolumn{3}{c|}{Utterance level}
        & \multicolumn{3}{c}{System level}
        \\
        & MSE & LCC & SRCC
        & MSE & LCC & SRCC
        & MSE & LCC & SRCC
        & MSE & LCC & SRCC
        \\
        \midrule
        magspec 
            & 0.4092 & 0.7668 & 0.7702
            & 0.1492 & 0.9080 & 0.9094
            & 0.4088 & 0.7833 & 0.7780
            & 0.2352 & 0.8608 & 0.8528
        \\
        melspec 
            & 0.3111 & 0.8034 & 0.8044
            & \textbf{0.1101} & 0.9235 & 0.9187
            & 0.3671 & 0.7820 & 0.7787
            & 0.2098 & 0.8610 & 0.8660
        \\
        wav2vec 
            & 0.4114 & 0.8222 & 0.8230
            & 0.1969 & 0.9329 & 0.9348
            & 0.4638 & 0.8226 & 0.8227
            & 0.2730 & 0.8936 & 0.8969
        \\
        \midrule
        \pbox{2.1cm}{wav2vec\\~+f0}
            & 0.3350 & 0.8320 & 0.8300
            & 0.1568 & 0.9340 & 0.9349
            & 0.3228 & \textbf{0.8395} & \textbf{0.8341}
            & 0.1641 & 0.9088 & \textbf{0.9084}
        \\
        \pbox{2.1cm}{wav2vec\\+melspec}
            & \textbf{0.3046} & \textbf{0.8414} & \textbf{0.8406}
            & 0.1321 & \textbf{0.9419} & \textbf{0.9417}
            & \textbf{0.3226} & 0.8346 & 0.8289
            & \textbf{0.1435} & \textbf{0.9097} & 0.9047
        \\
        \bottomrule
    \end{tabular}
    
    \hspace{10pt}
    \label{tab:performance}
\end{table*}

%% UMAP
\begin{figure*}[!ht]
    \centering
    \begin{minipage}[b]{0.3\linewidth}
        \centering
        \includegraphics[width=0.9\linewidth]{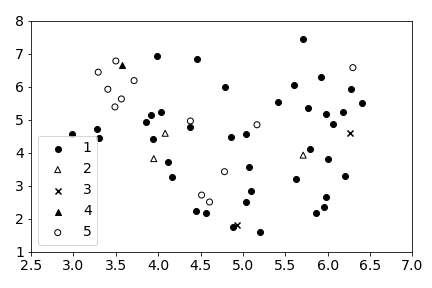}
        \\\vspace{-5pt}
        \textbf{magspec}
    \end{minipage}
\hfill
    \begin{minipage}[b]{0.3\linewidth}
        \centering
        \includegraphics[width=0.9\linewidth]{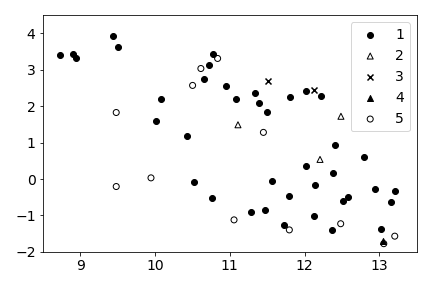}
        \\\vspace{-5pt}
        \textbf{melspec}
    \end{minipage}
\hfill
    \begin{minipage}[b]{0.3\linewidth}
        \centering
        \includegraphics[width=0.9\linewidth]{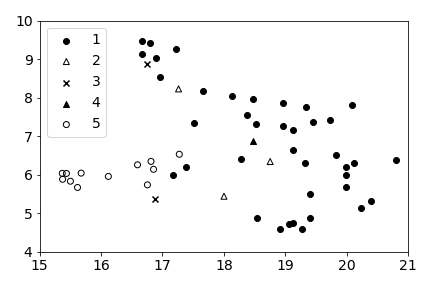}
        \\\vspace{-5pt} 
        \textbf{wav2vec}
    \end{minipage}
\hfill 
    \\(a) Training data

    \begin{minipage}[b]{0.3\linewidth}
        \centering
        \includegraphics[width=0.9\linewidth]{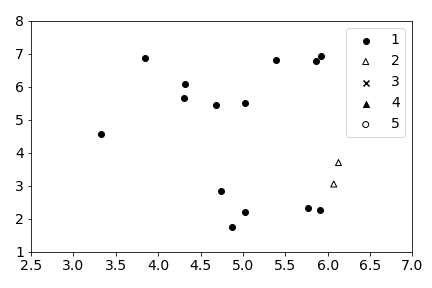}
        \\\vspace{-5pt}
        \textbf{magspec}
    \end{minipage}
\hfill
    \begin{minipage}[b]{0.3\linewidth}
        \centering
        \includegraphics[width=0.9\linewidth]{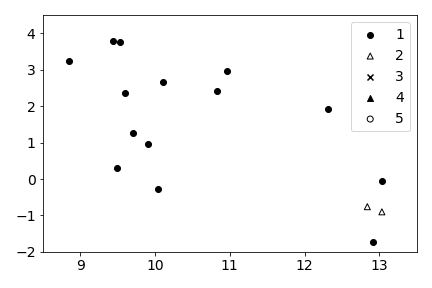}
        \\\vspace{-5pt}
        \textbf{melspec}
    \end{minipage}
\hfill
    \begin{minipage}[b]{0.3\linewidth}
        \centering
        \includegraphics[width=0.9\linewidth]{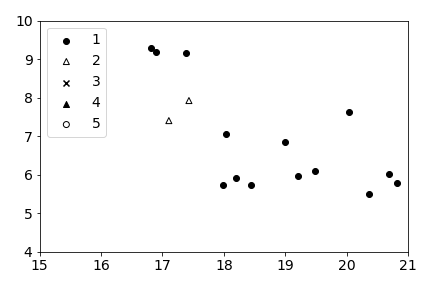}
        \\\vspace{-5pt}
        \textbf{wav2vec}
    \end{minipage}
\hfill
    \\(b) Test Data
\caption{The features on the UMAP plane, which was learned with all the samples that all 8 listeners assigned the same score.}
\label{fig:umap}
\vspace{-10pt}
\end{figure*}

\subsection{Dataset}
The data used in our experiments were provided by the VoiceMOS Challenge\footnote{\href{https://codalab.lisn.upsaclay.fr/competitions/695}{https://codalab.lisn.upsaclay.fr/competitions/695}}. 
The data is sampled from BVCC~\cite{bvcc}, the large-scale MOS dataset containing samples from the past BCs and VCCs. The train, valid and test sets include 4974, 1066, and 1066 audio samples (16kHz/16bit), respectively.
Each sample was rated by 8 listeners. In the valid and test sets, there were 8 listeners who were not included in the training set. 
%There are 7106 samples, with each sample rated by 8 listeners. In total there are 304 listeners, with each listener rating 187 samples. A carefully curated rule was used to create a 4974/1066/1066 train/valid/the test data split.
%There are 288 listeners in the training set, and there are 8 unseen listeners in the valid and test sets, with some overlap between the training listeners. 
For details, please refer to~\cite{sslmos}.

\subsection{Baseline system}
We used the official implementation of the LDNet\footnote[2]{\href{https://github.com/unilight/LDNet}{https://github.com/unilight/LDNet}}.
Its architecture was depicted in figure \ref{fig:ldnet}.
Amongst the three inference modes of LDNet, we chose the mean listener mode which showed the best system-level SRCC on the BVCC dataset \cite{ldnet}. The main differences between the experimental settings of the original LDNet paper and those of our experiments are described below.

\subsubsection{Loss function}
While the original LDNet used a Mean Square Error (MSE) as a loss function, we examined MSE, %negative Pearson Correlation Coefficient (PCC) and their combination: $MSE + k*PCC$ ($k$ is set to be 1 or 2) in a preliminary experiment. 
negative Linear Correlation Coefficient (LCC) and their combination: $MSE + k*LCC$ ($k$ is set to be 1 or 2) in a preliminary experiment. 
We found that LCC or $MSE + k*LCC$ outperformed MSE regarding SRCC. 
We chose therefore LCC as a loss function throughout the experiment in this paper. 
To calculate reliable LCC, a large batch size is crucial. 
We empirically found that 15 is sufficiently large and efficient in practice.

\subsubsection{Features}
While MOSNet~\cite{mosnet}, MBNet~\cite{mbnet} and LDNet~\cite{ldnet} used a 257-dimensional magnitude spectrogram (denoted as \textbf{magspec}) as the input feature, we investigated additional feature sets: melspectrogram and SSL-based feature, i.e. the latent representation of wav2vec.

For the task of MOS prediction, it would be beneficial to use a speech representation which more closely resembles human's perceptual scale. 
We extracted 80-dimensional melspectrogram (denoted as \textbf{melspec}) in a configuration of a 512-point-Short-Time-Fourier-Transform (STFT) with 128-point-shifting.

As Cooper et al. reported that speech representations by self-supervised learning (SSL) demonstrated good performance~\cite{sslmos}, we adopted wav2vec, one of the publicly available %self-supervised-learning-based (SSL) 
SSL-based speech models from the Fairseq\footnote{\href{https://github.com/pytorch/fairseq}{https://github.com/pytorch/fairseq}} project. 
Although the wav2vec 2.0~\cite{wav2vec2} produces the state-of-the-art performance~\cite{sslmos}, we adopt the 1.0 in  consideration of the smaller dimensions (512) and the size of the given data set.

Since the wav2vec architecture learned the abstract but essential representation of raw speech signals in a self-supervised manner, it is hard to assure if it captures or not some properties of raw speech signals that particularly relate to the target task (i.e. automatic MOS prediction in this work). 
Although the direct use of raw speech signals or adding magnitude of spectrogram might minimize missing essentials, learning will suffer from the curse of dimensionality.

On the other hand, melspectrogram is a relatively compressed representation but still retains many essentials that are useful across various speech applications (e.g. Automatic Speech Recognition, Text-To-Speech, etc.). 
F0 is even more compact than melspectrogram but carries out essential information about naturalness. 
Hence, we experiment the combination of the wav2vec and melspectrogram/F0 feature sets. 
Let us denote the combination of the wav2vec and melspectrogram and that of the wav2vec and F0 as \textbf{wav2vec+melspec} and \textbf{wav2vec+f0}, respectively. To build \textbf{wav2vec+melspec} feature, the wav2vec feature\footnote{For the detailed setting of the wav2vec feature extraction, please refer to the original paper \cite{wav2vec}} (512 x $n$) and melspectum (80 x $m$) are extracted from the raw audio as described above. 
Since the time dimensions ($n$ and $m$) are not equal to each other, melspectrogram is stretched to match $n$ by using a linear interpolation. 
The concatenation outputs the shape of (592 x n). 
In the same manner, 513-dimensional \textbf{wav2vec+f0} feature is built. For F0 extraction, pyworld~\footnote{\href{https://github.com/JeremyCCHsu/Python-Wrapper-for-World-Vocoder}{https://github.com/JeremyCCHsu/Python-Wrapper-for-World-Vocoder}} is used.

%%========================
%% Result and Discussion
%%========================
\section{Result and Discussion}\label{sec:result}
In this section, we present and discuss the results of our experiments. 
Firstly, we compare the performance of \textbf{magspec}, \textbf{melspec}, and \textbf{wav2vec}. 
Secondly, we investigate if the performance improves by adding melspectrogram/F0 features (~\textbf{wav2vec+melspec} and ~\textbf{wav2vec+f0}~). 
Lastly, we will discuss the challenges and the future research direction.

%%================================
%% magspec vs melspec vs wav2vec
%%================================
\subsection{magspec vs. melspec. vs. wav2vec}\label{sec:comparison}
Each model was trained until 200k steps, but the best model was chosen by monitoring system-level SRCC on the validation data set. 
We report not only SRCC but also LCC and MSE as a reference. 
Note that we did not optimize models by %LCC or 
MSE.

The results are shown in Table \ref{tab:performance}. 
In all conditions (~i.e., Utterance-level or System-level or with the validation data or with the test data~), \textbf{wav2vec} performed the best regarding SRCC. 
The second place was \textbf{melspec} while \textbf{magspec} performed the worst. 
As reported in the previous study~\cite{sslmos}, \textbf{wav2vec} showed the strong generalization capability. 

To have a better understanding of the representations, we visualize them by performing Uniform Manifold Approximation and Projection (UMAP)~\cite{umap}. 
UMAP is learned with the entire dataset i.e. training + validation + test set. 
To filter out unreliable MOS scores, we chose samples that all 8 listeners assigned the same score. 
Firstly, we see how discriminating they are in the training materials. 
Next, we see how the discrimination are well retained in the test data, which shows the generalization capability. 

Figure \ref{fig:umap} (a) shows the results on the the training data. 
Both \textbf{magspec} and \textbf{melspec} did not show clear clusters of each score but a scattered distribution. \textbf{wav2vec} resulted in more clear cluster for score 5 while score 1 is still dominant and scattered in the filtered data. 
Figure \ref{fig:umap} (b) shows the results on testing data. 
When we filtered out the samples that do not have a complete agreement between listeners, only samples with low scores (1 and 2) were left, which means that the majority of samples was more ambiguous to rate.
Due to the restriction of filtering, we could not find any clear clusters in all feature sets. 
Since \textbf{wav2vec} performed the best on the test data in spite of the challenge posed by the ambiguity, our experiment firms the previous finding: the strong generalization of \textbf{wav2vec}.

%%================================
%% Combination of representations
%%================================
\subsection{Combinations of representations}\label{sec:combination}
As described in Section \ref{sec:experiment}, we are not sure if \textbf{wav2vec} retains enough properties of raw speech signals that are relevant to the automatic MOS prediction task. 
In this section, we compare \textbf{wav2vec},  \textbf{wav2vec+melspec}, and \textbf{wav2vec+f0} as shown in Table~\ref{tab:performance}. 
When melspectrogram or F0 feature is added to the wav2vec features, both MSE and SRCC improved. 
In validation set, \textbf{wav2vec+melspec} surpasses the performance of \textbf{wav2vec+f0} while \textbf{wav2vec+f0} shows slightly higher SRCC in test set. 

\begin{figure}[t]
    \centering
    \includegraphics[width=\linewidth]{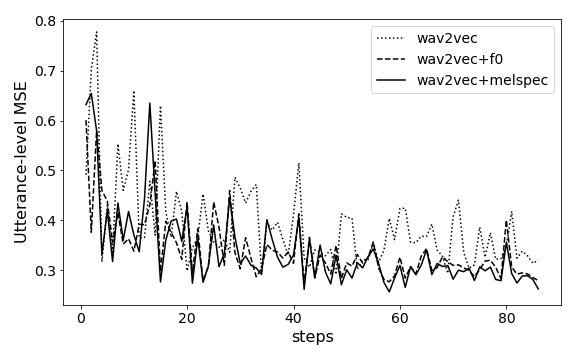}
    \vspace{-10pt}
    \caption{Learning curve of Utterance-level MSE}
    \label{fig:utt_mse}
    \vspace{-10pt}
\end{figure}

Based on this finding, we question where these gains come from. We investigated how the addition helps learning and generalization. 
Firstly, we observed that the loss changes have a smaller variation than those without the additional features as shown in Figure~\ref{fig:utt_mse}. 
Hence, we conclude that both melspectrogram and F0 feature lead to more smooth optimization. 
%% learning curve

\begin{figure}[t]
    \centering
    \includegraphics[width=0.9\linewidth]{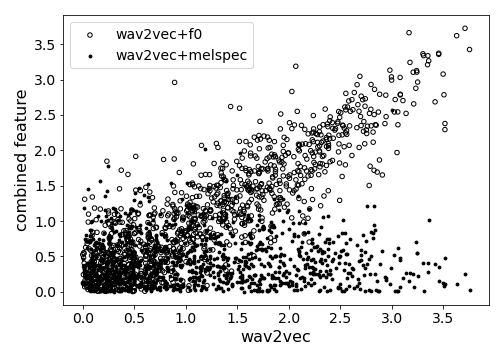}
    \vspace{-10pt}
    \caption{The absolute deviation between the predicted score and the mean score i.e. $| \bar{s}_i - f(x_i) |$ of the test data.}
    \vspace{-10pt}
    \label{fig:mos_dif}
\end{figure}

Secondly, we visualized the absolute deviation i.e. $| \bar{s}_i - f(x_i) |$ of combined features against \textbf{wav2vec} in the test data. 
As shown Figure~\ref{fig:mos_dif}, the absolute deviation of \textbf{wav2vec+f0} has the high correlation to that of \text{wav2vec} (LCC of 0.875). 
In other words, to the samples of which \textbf{wav2vec} performs well \textbf{wav2vec+f0} also performs well, and when \textbf{wav2vec} produces errors \textbf{wav2vec+f0} produces errors too. 

There was not much difference between \textbf{wav2vec} and \textbf{wav2vec+melspec} when the absolute deviation is below 1. 
However, when it becomes large, \textbf{wav2vec+melspec} greatly outperforms \textbf{wav2vec}.
Since \textbf{wav2vec+melspec} well predicted the samples that \textbf{wav2vec} did not, melspectrogram seems to retain some essentials that \textbf{wav2vec} might fail to capture.
%% diff.

%%=============
%% Discussion
%%=============
\subsection{Discussion}\label{sec:discussion}
The results above imply that melspectrogram and F0 feature sets retain the important clue that the wav2vec features might leave out. Regardless, some samples are always erroneous. 
Intuitively, samples with less reliable scores are more challenging to learn and may hurt learning. 
We attempted to quantify what makes listeners less agree with each other but could not find significant clues. 
Rather, we assume that listeners rely on various aspects of the quality and have different weights. 
For instance, one listener weighs more accuracy of prosody than that of pronunciation while another focuses on fidelity of audio (mostly affected by vocoder). 
Filtering out unreliable samples might improve learning; regardless, the system may encounter ill-defined samples in the wild since that might be the nature of subjective scores. 
Let us leave this issue as one of future studies.

Lastly, we did not conduct experiments with out-of-domain data sets that will make generalization even harder. 
However, due to the large scale but the aggregated dataset from various contexts, the generalization would be still challenging in this study. We will investigate the challenge of out-of-domain data sets in the future as well.

%%==============================
%% Conclusions and Future Work
%%==============================
\section{Conclusions and Future Work}\label{sec:conclusion}
In this work, we investigated the generalization of speech representations including magnitude of spectrogram, melspectrogram, F0, and wav2vec. 
We found the wav2vec feature outperformed others on the test data where the reliability of scores was often low. 
In addition, the combination of the wav2vec and melspectrogram or F0 feature sets further improved the performance. 
This finding tells us that the wav2vec feature set has missing information and the combinations fills the gap.

To validate our findings in a more rigorous setup, we will experiment them with an out-of-domain dataset or a cross-corpora setup that poses a great difficulty of generalization. 
Moreover, we will investigate how to leverage the use of the variation of ratings from multiple listeners that might give the clues of ill-defined data.

%%===================
%% Acknowledgements
%%===================
\section{Acknowledgements}\label{sec:ack}
The authors would like to thank the organizers of the VoiceMOS Challenges for providing the data set.

\newpage
\bibliographystyle{IEEEtran}
\bibliography{bibliography}

\end{document}